\documentclass[a4paper,12pt]{article}

\usepackage[a4paper]{geometry}
\usepackage{amsmath}
\usepackage{amssymb}
\usepackage{listings}
\usepackage[numbers]{natbib}
\usepackage[T1]{fontenc}
\usepackage[utf8]{inputenc}
\usepackage{authblk}
\usepackage{color,graphicx}
\usepackage{array}
\usepackage{setspace}
\usepackage{siunitx}
\usepackage{pgfplots}
\pgfplotsset{compat=newest}
\usepackage{tikz}
\usetikzlibrary{arrows,shapes,chains, decorations.markings}
\usepackage{aicescover}
\usepackage{hyperref}

\definecolor{myred}{RGB}{204, 0, 0} 
\definecolor{myred2}{RGB}{242, 122, 145} 
\definecolor{myblue}{RGB}{48, 145, 242} 
\definecolor{myorange}{RGB}{242, 139, 36} 

\everymath{\displaystyle}

\begin{document}
\title{A Note on Time Measurements in LAMMPS}
\author[1]{Daniel Tameling\thanks{tameling@aices.rwth-aachen.de}}
\author[1]{Paolo Bientinesi}
\author[1,2]{Ahmed E. Ismail}
\affil[1]{AICES Graduate School, RWTH Aachen University, Schinkelstr.\ 2, 52062 Aachen, Germany}
\affil[2]{Aachener Verfahrenstechnik: Molecular Simulations and Transformations, Faculty of Mechanical Engineering, RWTH Aachen University, Schinkelstr.\ 2, 52062 Aachen, Germany}
\date{\today}
\aicescoverauthor{Daniel Tameling, Paolo Bientinesi, and Ahmed E. Ismail}
\aicescoverpage
\maketitle
\begin{abstract} We examine the issue of assessing the efficiency of
  components of a parallel program at the example of the MD package
  LAMMPS. In particular, we look at how LAMMPS deals with the issue
  and explain why the approach adopted might lead to inaccurate conclusions. The
  misleading nature of this approach is subsequently verified
  experimentally with a case study. Afterwards, we demonstrate how one
  should correctly determine the efficiency of the components and show
  what changes to the code base of LAMMPS are necessary in order to
  get the correct behavior.
\end{abstract}

One of the governing themes of our times is the efficient use of
resources. The rule of this theme is not just limited to natural
resources like oil but it extends as well to computational resources.
In particular, the efficient usage of computational resources
constitutes a pressing issue for extremely demanding applications like
molecular dynamics (MD) simulations. However, in contrast to the
natural resources, the main concern here is not to go easy on the
resources; it is feasibility. To be efficient can make in this context
the difference between getting the results of a simulation towards the
end of the century and getting them within a few weeks. So to increase
the efficiency can be the key to make it possible at all to carry out
crucial research. Even if one is in the lucky situation that the
simulation takes only a few weeks, any substantial improvement in its
efficiency will save a significant amount of simulation time. But in
order to improve the efficiency of such a simulation, it is vital to
know how much time is consumed by each of its elements. Thus, we will
discuss in this note how timings are measured by the MD package
LAMMPS~\cite{Plimpton1995}, and why the presentation of these
measurements can constitute an inaccurate representation of the
efficiency of the executed components.

The main purpose of this note is to rise the awareness in the MD
community for how timings are measured correctly in parallel
applications. While the discussed issues are relevant for every
discipline that uses parallel applications, we have the feeling that
in this particular community there exists a lack of the knowledge
about the discussed matter, which is the reason why LAMMPS outputs
misleading information. Despite the fact that we pick LAMMPS as an
example, the unfolding discussion is equally relevant for the authors
and users of other MD packages, such as GROMACS~\cite{Hess2008}, and
NAMD~\cite{Kale1999}. Moreover, because of the nature of the discussed
matter, it is the case the presented recommendations are also valuable
for researchers in other fields if they rely on parallel programs, and
we want to stress that this note does not require any prerequisite
knowledge about MD.

The challenge that this note is going to address is the difficulty of
producing reliable and at the same time easily understandable
information about the efficiency of parts of a parallel program. In
case of a serial application, this task is trivial: the efficiency of
a part is directly correlated to the wall-clock time consumed by this
component. But in a parallel application, there exists such a timing
for each process that in general differ from each other. The question
is now how to extract representative information about the efficiency
of the respective component from this potentially vast amount of data.
Ideally, each efficiency should be represented reliably by a single
number. Therefore, we will discuss here what value is suitable for
assessing the efficiency of the different elements of a parallel
application. Ideally, each element would have for this purpose just a
single number, which is exactly what LAMMPS attempts. However, as we
are going to show, the methodology that LAMMPS uses to produce this
number is inappropriate and can be misleading.

Before we take a closer look at LAMMPS, there are some questions that
need to be answered. In particular, these questions are why the MD
community uses timings as a measure for the efficiency of their
programs and why are the timings of parts of a program relevant at
all. The answer to the first question is that the efficiency of a MD
program is usually simply unknown. The traditional definition of the
efficiency of a computer program is the number of executed flops
divided by the number of flops theoretically possible within the time
spanned by the simulation. However, as MD programs constitute complex
applications, the amount of executed flops remains typically unknown.
Moreover, users of MD applications are normally not interested in the
efficiency of a program; they only care how long they have to wait
until they get a solution. Therefore, they commonly use as metric for
the efficiency of their program the time to solution.

Meanwhile, instead of the time to solution, the MD community often
relies on a different quantity that is normally completely equivalent:
the average time per step. The reason for this is that a typical MD
simulation consists of a large number of steps. Moreover, it is the
case that the in real applications the time spent for initialization
and finalization is negligible. Accordingly, the average time per step
times the number of steps equals approximately the time to solution.

After answering the first question, we proceed with addressing the
other one. In particular, we are going to disclose why the knowledge
of the timings of parts of a program constitutes a valuable
information. The first reason is that these timings indicate which
parts are the most promising candidates for optimization attempts.
Furthermore, to know that a certain component takes longer than a
specific amount of time can mean that it would be faster to employ an
alternative method to fulfill the given task. Finally, the knowledge
of how much time a certain method takes can be a vital ingredient for
the task of selecting its parameters optimally. Therefore, it is
immensely valuable to be aware of how long the execution of crucial
components of an application takes. At the same time, this means that
it can be extremely harmful to misrepresent the execution time of
these elements, like LAMMPS does.

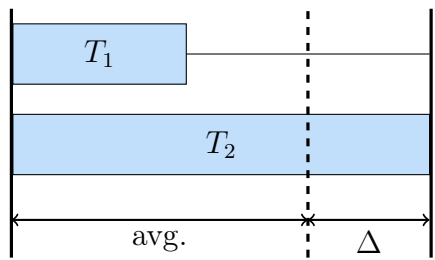
\begin{figure}[ht]
  \begin{center}
    \begin{tikzpicture}[start chain=going right, node distance=-0.1]
      \draw [very thick] (0,-1.1) -- (0,2.2);
      \draw [fill=myblue!30] (0.02,1.2) rectangle (2.3,2) node[pos=.5] {$T_{1}$};
      \draw [fill=myblue!30] (0.02,0) rectangle (5.5,.8) node[pos=.5] {$T_{2}$};
      \draw (2.3,1.6) -- (5.5,1.6);
      \draw [very thick] (5.52,-1.1) -- (5.52,2.2);
      \draw [very thick, dashed] (3.9,-1.1) -- (3.9,2.2);
      \draw [<->, thick] (0,-.6) -- (3.9,-.6) node[pos=.5, below] {avg.};
      \draw [<->, thick] (3.9,-.6) -- (5.5,-.6) node[pos=.5, below] {$\Delta$};
    \end{tikzpicture}
    \caption{The average across all ranks for a function consisting of
    one part is smaller than the actual run time, which equals the maximum.}
    \label{fig:avg-erroneous}
  \end{center}
\end{figure}

The reason why LAMMPS produces misleading information is that the
package always averages the timings across the individual processes. To
illustrate why the average is problematic, we will review its effects
for a simple example: a parallel program with $n$ processes that
consists of just one component that is executed for a single time
step. This example is depicted graphically in
\autoref{fig:avg-erroneous}. The program terminates as soon as all
processes complete; correspondingly, the execution time
$T_\text{exec}$ equals the maximum of the individual timings
$T_\text{max}$. One would expect that the reported timing for the only
component would equal this maximum. But if one puts the average of all
the timings in relation to the execution time, one obtains
\begin{equation}
  T_\text{avg} = \frac{1}{n} \sum_{i=1}^n T_i
  \le \frac{1}{n} \sum_{i=1}^n T_\text{max}
  = \frac{1}{n} \cdot n \cdot  T_\text{max}
  = T_\text{max} = T_\text{exec}.
  \label{eq:avg-smaller-than-max}
\end{equation}
Here one has to realize that the equality holds only when all
processes take the same amount of time. But for MD simulations it is
highly unlikely that all of the individual timings would exhibit the
exact same value. Therefore, the average of the timings would be
misleading; it would be lower than the real time consumption of the
respective component.

How treacherous the average potentially is can be demonstrated by
looking at how it behaves in scaling experiments of the example
program. Meanwhile, in order to judge the results of the average, one
needs something to compare them to. For this, we use the best possible
scaling. Moreover, we introduce the assumption that the
parallelization introduced no overhead. Consequently, in the optimal
case, the execution time on $n$ processes will equal
\begin{equation}
  T_\text{exec}^n = \frac{T_0}{n},
  \label{eq:parallel-execution}
\end{equation}
where $T_0$ denotes the serial run time. One has to notice that this
optimal reduction occurs only if the work is equally distributed among
the processes; if the individual timings of the involved processes are
not exactly the same, the execution time will be larger than indicated
by this equation.

Now we look at how the average of the individual timings behaves. For
this, we exploit that there is no overhead associated with the
parallelization. Consequently, it is the case that the sum of the
individual timings equals the serial time:
\begin{equation}
  T_0 = \sum_{i=1}^n T_i.
  \label{eq:T0-equal-sum}
\end{equation}
This relation can now be used to calculate the average of the
individual timings $T_i$:
\begin{equation}
  T_\text{avg} = \frac{1}{n} \sum_{i=1}^n T_i
  = \frac{T_0}{n},
  \label{eq:avg-similar-to-optimal}
\end{equation}
which surprisingly equals the expression of
\autoref{eq:parallel-execution}, which signifies the optimal time
reduction. But here, we never assumed anything about the distribution
of the timings. In particular, we did not introduce that the work is
equally distributed. Therefore, no matter how imbalanced the workload
is in reality, considering the average of the timings will always
indicate the optimal scaling. But this is not the case. As we have
shown above, the run time of our example program is actually the
maximum of the individual timings. Therefore, the average of the
timings can mislead the user regarding the scaling of the program. By
which degree depends totally on the distribution of the workload. In
fact, using an extremely poor parallelization strategy, such as having
one process execute the serial program while the other $n - 1$
processes remain idle, we would have optimal speedup if we consider
the average, while the actual run time of the program would not
improve at all.

Meanwhile, it might be the case that in realistic MD simulations all
the components are always perfectly load balanced. Even if that is not
the case, maybe the deviations of the average from the correct result
are so small that they are negligible. Therefore, it is now the time
to put the theoretical considerations to the test. To do this, we will
use two different test systems. The first has an homogeneous and the
second an inhomogeneous particle distribution. Correspondingly, there
should be no load imbalance, except for small statistical variations
that occur in typical atomic systems. But this case constitutes the
best case scenario; the load balancing with any realistic system will
not be better than this. Hence, no matter how large the deviation of
the average from the correct result will be in this scenario, it will
never be significantly smaller. In contrast, the second system does
not exemplify the worst case scenario. Sometimes one has to live with
even larger inhomogeneities in order to be able to simulate the
system.

For the component whose timings we are going to investigate we picked
a standard task that is part of basically every MD
simulation~\cite{Frenkel2001,Griebel2007}. Specifically, the component
will concentrate on is the force calculation with a cutoff, which in
LAMMPS is reported under the label ``Pair''. Moreover, we kept the
physical nature systems as simple as possible. Accordingly, we set up
the system as a Lennard-Jones fluid to which we applied a cutoff of
$3\sigma$. Finally, we want to mention that despite the simplicity of
the system, the simulation consists of more components than the
cutoff-based computation. Nevertheless, in the following we consider
only the timings of this component.

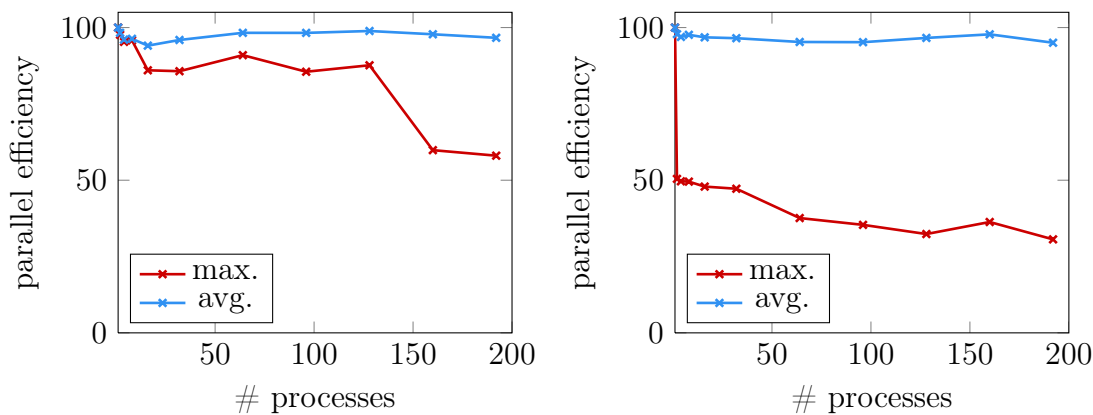
\begin{figure}[ht]
  \begin{center}
    \pgfplotsset{every axis plot/.append style={line width=1pt}}
    \begin{tikzpicture}
      \begin{axis}[width=0.46\textwidth, xlabel=\# processes, ylabel=
        parallel efficiency, ymin=0, ymax=105, xmin=1, xmax=200, legend pos=south west]

        \addplot[color=myred,mark=x]
        table[x=procs, y expr=100*\thisrow{maximum}] {scaling_hom.txt};
        \addplot[color=myblue,mark=x]
        table[x=procs, y expr=100*\thisrow{average}] {scaling_hom.txt};
        
        \legend{max., avg.}
      \end{axis}
    \end{tikzpicture}
    \begin{tikzpicture}
      \begin{axis}[width=0.46\textwidth, xlabel=\# processes, ylabel=
        parallel efficiency, ymin=0, ymax=105, xmin=1, xmax=200, legend pos=south west]

        \addplot[color=myred,mark=x]
        table[x=procs, y expr=100*\thisrow{maximum}] {scaling_inhom.txt};
        \addplot[color=myblue,mark=x]
        table[x=procs, y expr=100*\thisrow{average}] {scaling_inhom.txt};
        
        \legend{max., avg.}
      \end{axis}
    \end{tikzpicture}
    \caption{Parallel efficiency as indicated by the average of the
      individual timings and their maximum. Left: Homogeneous system.
      Right: Inhomogenous system.}
    \label{fig:comparison-scaling-max-avg}
  \end{center}
\end{figure}

First of all, we look at the results of the best case scenario, i.e.
the homogeneous system. This system was created by filling the whole
domain with particles. As it is the best case scenario, the difference
between the maximum and the average should be as small as it can
become in practice. This is what makes the results that are 
depicted in the left plot of
\autoref{fig:comparison-scaling-max-avg}
particularly disappointing. Even for a relatively low number of
processes, one can observe that it makes a significant
difference whether one considers the average or the maximum. Moreover,
the disagreement becomes worse as the number of processes is
increased. The conclusion is that even under optimal circumstances the
usage of the average to assess the performance of a part of the
simulation is inadequate.

Nevertheless, it is also important to be aware of how bad the
disagreement can become. Therefore, we discuss in this paragraph the
results for the inhomogenous system. Here it should be noted that we
created the inhomogeneity so that in all simulations the particles are
assigned to only half of the employed processes. The consequence of
this setup is that during the considered cutoff-computation half of
the processes are idle. Thus, one can expect that the parallel
efficiency is at best at 50\%. As one can see in the right plot of
\autoref{fig:comparison-scaling-max-avg}, this expectation is
fulfilled by the maximum of the individual timings: for any parallel
runs, it measures an efficiency below 50\%. By contrast, the average
gives a completely different impression: it looks like the scaling is
almost perfect. The transported message constitutes just a gross
inconsistency with reality. Moreover, as mentioned, there are even
worse inhomogeneities in realistic systems. Thus, given the right
circumstances, the corresponding prediction of the average will become
arbitrarily wrong. Together with its behavior in the homogeneous
experiments, the overall verdict for the average can only be that it
is totally inappropriate for assessing the performance of components
of parallel application.

\begin{figure}[ht]
  \begin{center}
    \begin{tikzpicture}[start chain=going right, node distance=-0.1]
      \draw [very thick] (0,-.2) -- (0,2.2);
      \draw [fill=myblue!30] (0.02,0) rectangle (2.5,.8) node[pos=.5] {$T_{21}$};
      \draw [fill=myblue!30] (0.02,1.2) rectangle (1.3,2) node[pos=.5] {$T_{11}$};
      \draw (3.3,1.6) -- (3.5,1.6);
      \draw [fill=myred2!30] (2.5,0) rectangle (3.5,.8) node[pos=.5] {$T_{22}$};
      \draw [fill=myred2!30] (1.3,1.2) rectangle (3.3,2)node[pos=.5] {$T_{12}$};
      \draw [very thick] (3.52,-.2) -- (3.52,2.2);
    \end{tikzpicture}
    \hspace*{2cm}
    \begin{tikzpicture}[start chain=going right, node distance=-0.1]
      \draw [very thick] (0,-.2) -- (0,2.2);
      \draw [fill=myblue!30] (0.02,0) rectangle (2.5,.8) node[pos=.5] {$T_{21}$};
      \draw [fill=myblue!30] (0.02,1.2) rectangle (1.3,2) node[pos=.5] {$T_{11}$};
      \draw (1.3,1.6) -- (2.5,1.6);
      \draw [very thick] (2.52,-.2) -- (2.52,2.2);
      \draw [fill=myred2!30] (2.54,0) rectangle (3.5,.8) node[pos=.5] {$T_{22}$};
      \draw [fill=myred2!30] (2.54,1.2) rectangle (4.5,2) node[pos=.5] {$T_{12}$};
      \draw (3.5,.4) -- (4.5,.4);
      \draw [very thick] (4.52,-.2) -- (4.52,2.2);
    \end{tikzpicture}
    \caption{Barriers help to separate individual tasks, but usually will lead to a reduction in performance.}
    \label{fig:timing-barrier}
  \end{center}
\end{figure}
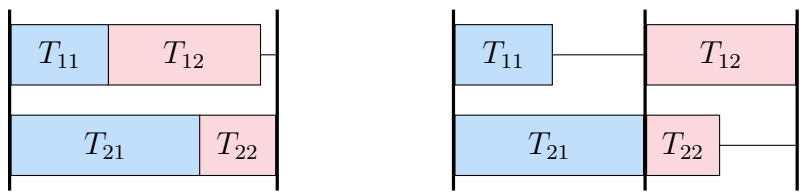

But what are the alternatives to using the average? Well, one solution
is hidden in the source code of LAMMPS. The function that is
responsible for the timings contains an interesting comment:
\begin{lstlisting}
  void Timer::stamp(int which)
  {
    // uncomment if want synchronized timing
    // MPI_Barrier(world);
    double current_time = MPI_Wtime();
    array[which] += current_time - previous_time;
    previous_time = current_time;
  }
\end{lstlisting}
What happens if the comment is removed? Every element whose time is
measured will start simultaneously across all processes, and even if
one process finishes the corresponding tasks, it will wait until all
of them completed these tasks. Thus, all the processes will measure
the time that the slowest process took. Accordingly, as all processes
spent the same time in every component, the average the individual
timings produces a correct assessment of the corresponding
performance. (Meanwhile, one could also just report the timings of an
arbitrary process; the result would be the same.) However, it has its
reasons why the barrier is commented; its usage comes at a price. As
illustrated in \autoref{fig:timing-barrier}, the barrier can namely
reduce the performance. This would be an issue if the goal is to
assess an element that extends across multiple components as it is,
for example, the case for a long-range solver. Furthermore, since
typical MD simulations last from days to weeks, even the tiniest
slowdown is intolerable. In conclusion, because it might harm the
performance of the program, it is not recommended to introduce a
barrier between its different components.

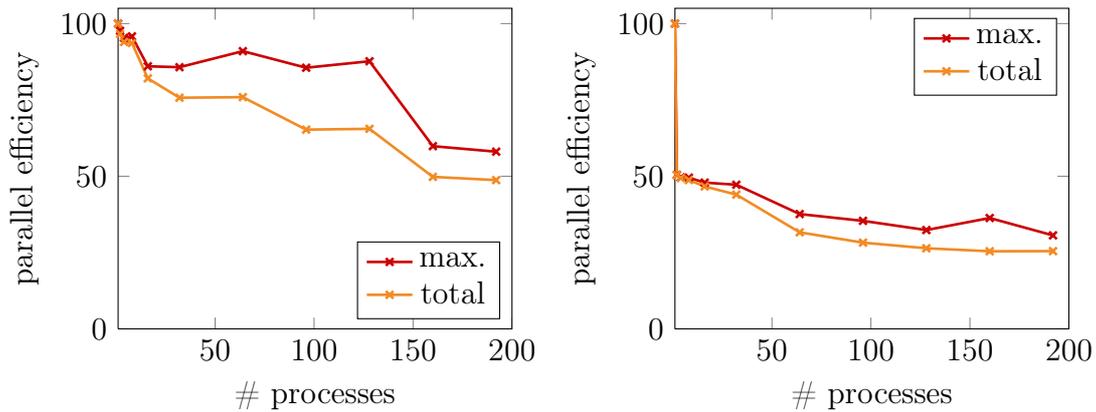
\begin{figure}[ht]
  \begin{center}
    \pgfplotsset{every axis plot/.append style={line width=1pt}}
    \begin{tikzpicture}
      \begin{axis}[width=0.46\textwidth, xlabel=\# processes, ylabel=
        parallel efficiency, ymin=0, ymax=105, xmin=1, xmax=200, legend pos=south east]

        \addplot[color=myred,mark=x]
        table[x=procs, y expr=100*\thisrow{maximum}] {scaling_hom.txt};
        \addplot[color=myorange,mark=x]
        table[x=procs, y expr=100*\thisrow{total}] {scaling_hom.txt};
        
        \legend{max., total}
      \end{axis}
    \end{tikzpicture}
    \begin{tikzpicture}
      \begin{axis}[width=0.46\textwidth, xlabel=\# processes, ylabel=
        parallel efficiency, ymin=0, ymax=105, xmin=1, xmax=200, legend pos=north east]

        \addplot[color=myred,mark=x]
        table[x=procs, y expr=100*\thisrow{maximum}] {scaling_inhom.txt};
        \addplot[color=myorange,mark=x]
        table[x=procs, y expr=100*\thisrow{total}] {scaling_inhom.txt};
        
        \legend{max., total}
      \end{axis}
    \end{tikzpicture}
    \caption{Parallel efficiency as indicated by the maximum of the
      individual timings and the duration of a complete step maximum.
      Left: Homogeneous system. Right: Inhomogenous system.}
    \label{fig:comparison-scaling-max-total}
  \end{center}
\end{figure}

Is there a better solution than the barrier? Yes, there is: to use the
maximum of the timings. In the example program that was discussed in
the beginning, the maximum provided always a correct assessment of the
performance. Moreover, in the conducted experiments, the use of the
maximum resulted in predictions of the efficiency that were in
agreement with the general expectations. In fact, the comparison in
\autoref{fig:comparison-scaling-max-total} reveals that the maximum
behaves similarly as the total time to solution. This is again a sign
for the quality of the maximum since it is the considered
cutoff-computation that dominates the cost of the simulation. To sum
up, the maximum of the timings is the appropriate measure for the
performance.

To use the maximum instead of the average requires only very small
changes to the code base of LAMMPS. Specifically, one has to modify a
few lines in the \texttt{finish.cpp} file. Specifically, whenever the
average of timings is calculated as in
\begin{lstlisting}
// default version with average of timings
time = timer->array[TIME_PAIR];
MPI_Allreduce(&time,&tmp,1,MPI_DOUBLE,MPI_SUM,world);
time = tmp/nprocs;
\end{lstlisting}
the following modifications are sufficient
\begin{lstlisting}
// changed version with maximum of timings
time = timer->array[TIME_PAIR];
MPI_Allreduce(&time,&tmp,1,MPI_DOUBLE,MPI_MAX,world);
\end{lstlisting}
Basically, one has to replace the \texttt{MPI\_SUM} in the
corresponding \texttt{allreduce} calls by an \texttt{MPI\_MAX}, and
one has to remove the divisions by the number of processes.

If one uses the maximum of the timings, there is one thing one should
be aware of: the sum of the maxima of all components will in genera;
be larger than the total execution time. This behavior can be observed
in \autoref{fig:timing-barrier}: the sum of the maxima is
$T_{21}+T_{12}$, whereas the execution equals $T_{21}+T_{22}$. One
consequence of this phenomenon is always present in addition to the
timings of the components also the total execution time. There is
another reason why presenting the total execution time is important.
It is namely also relevant for comparing different timings of a single
component. For example, for the case depicted in
\autoref{fig:timing-barrier}, one could optimize the code and reduce
$T_{12}$ to one half of its original value. As the maximum value of
the timings of the second component of the program would also been cut
in half, one could claim that the optimization was extremely
successful. However, the overall execution time would remain the same.
Therefore, the optimization achieved no real benefit, which
exemplifies why it is important to discuss the total execution time.

In this note we explored the difficulties associated with assessing
the timings of elements of parallel applications at the example of the
MD package LAMMPS. We demonstrated the problems with the methodology
that is currently employed by default in this package and how to
resolve the existing problems. Specifically, for a realistic
assessment of the time consumption of a component, it is inevitable to
use the maximum of the corresponding timings across the corresponding
processes.

\bibliographystyle{abbrv}
\bibliography{literature}

\end{document}